\documentclass[twoside]{article}
\usepackage{fleqn,espcrc2}
\usepackage[psamsfonts]{amsfonts}

\def\rmd{{\rm d}}
\def\rmD{{\rm D}}
\def\rme{{\rm e}}
\def\rmO{{\rm O}}
\def\rmU{{\rm U}}

\def\rz{{\mathbb R}}

\def\Im{{\rm Im}\,}

\def\proof{\noindent{\sl Proof:}\kern0.6em}

\def\frac#1#2{\hbox{$#1\over#2$}}
\def\dual{\mathstrut^*\kern-0.1em}

\def\lvec#1{\setbox0=\hbox{$#1$}
    \setbox1=\hbox{$\scriptstyle\leftarrow$}
    #1\kern-\wd0\smash{
    \raise\ht0\hbox{$\raise1pt\hbox{$\scriptstyle\leftarrow$}$}}
    \kern-\wd1\kern\wd0}
\def\rvec#1{\setbox0=\hbox{$#1$}
    \setbox1=\hbox{$\scriptstyle\rightarrow$}
    #1\kern-\wd0\smash{
    \raise\ht0\hbox{$\raise1pt\hbox{$\scriptstyle\rightarrow$}$}}
    \kern-\wd1\kern\wd0}
\def\boxit#1{\vbox{\hrule height2pt\hbox{\vrule width2pt
    \kern10pt\vbox{\kern10pt#1\kern10pt}\kern10pt\vrule width2pt}
    \hrule height2pt}}

\def\nab#1{{\nabla_{\kern-1.0pt#1}}}
\def\nabstar#1{\nabla\kern-0.0pt\smash{\raise 4.5pt\hbox{$\ast$}}
               _{\kern-5.5pt#1}\kern0.5pt}

\def\drvstar#1{\partial\kern-0.0pt\smash{\raise 4.0pt\hbox{$\ast$}}
               _{\kern-5.5pt#1}\kern0.5pt}

\def\ldrvstar#1{\lvec{\,\partial}\kern1.0pt\smash{\raise 8.0pt\hbox{$\ast$}}
               \kern-13.0pt_{#1}\kern2.0pt}

\def\psibar{\overline{\psi}}
\def\anomaly{{\cal A}}

\def\dirac#1{\gamma_{#1}}
\def\diracstar#1#2{
    \setbox0=\hbox{$\gamma$}\setbox1=\hbox{$\gamma_{#1}$}
    \gamma_{#1}\kern-\wd1\kern\wd0
    \smash{\raise4.5pt\hbox{$\scriptstyle#2$}}}
\def\dirachat{\hat{\gamma}_5}

\def\tr{{\rm tr}}
\def\Tr{{\rm Tr}}

\def\L{{\mathfrak L}}

\def\trans{{\cal Q}}

\def\D{D^A}

\def\LxRtwo{\hbox{$\mathbb L${\kern0.70pt}attice}\times\rz^2}

\title{\vspace{-4.0cm}
       \rightline{\normalsize CERN-TH/99-290}
       \vspace{-0.1cm}
       \rightline{\normalsize September 1999}
       \vspace{2.9cm}
       Chiral gauge theories on the lattice with exact gauge invariance%
       \thanks{Talk given at
       the International Symposium on
       Lattice Field Theory, June 29 -- July 3, 1999, 
       Pisa, Italy}}

\author{M.~L\"uscher\,\address{Deutsches 
        Elektronen-Synchrotron DESY,
        Notkestrasse 85, D-22603 Hamburg, Germany}%
        $^{,}$%
        \address{CERN, Theory Division, CH-1211 Geneva 23, Switzerland
        (present address)}}

\begin{document}

\begin{abstract}
A recently proposed formulation of chiral lattice gauge theories
is reviewed, in which the locality and gauge invariance of the 
theory can be preserved
if the fermion representation of the gauge group is anomaly-free.
\end{abstract}

\maketitle

\section{INTRODUCTION}

At the quantum level chiral gauge theories 
have always been relatively difficult
to treat, because of the gauge anomaly 
and the related fact that
there seemed to be no way to regularize these theories
without breaking the gauge symmetry.
The famous no-go theorem of Nielsen and Ni\-no\-miya 
\cite{NN,Friedan} was often taken as a proof of this
and many people working on the subject suspected that 
the problem might reflect a fundamental limitation
of quantum field theory.
As a consequence it remained unclear 
whether anomaly-free chiral gauge theories
can be defined consistently
beyond perturbation theory.
The existence of global anomalies \cite{Witten,ElitzurNair}
in fact suggests that the answer
might be ``no" in some cases.

In this review we consider left-handed fermions,
\begin{equation}
  P_{-}\psi=\psi,
  \quad
  \psibar P_{+}=\psibar,
  \quad
  P_{\pm}\equiv\frac{1}{2}(1\pm\dirac{5}),
\end{equation}
transforming according to some unitary representation $R$ of the 
gauge group. The action density of the continuum 
theory is assumed to be of the standard form
\begin{equation}
  {\cal L}(x)={1\over4g^2}F_{\mu\nu}^a(x)F_{\mu\nu}^a(x)
  +\psibar(x)\dirac{\mu}D_{\mu}\psi(x),
\end{equation}
where $F^a_{\mu\nu}$ denotes the gauge field tensor and
$D_{\mu}$ the appropriate covariant derivative.
The gauge anomaly in this theory is proportional to
\begin{equation}
  d_R^{abc}\equiv2i\,\tr\{R(T^a)R(T^b)R(T^c)+(b\leftrightarrow c)\}
\end{equation}
with $R(T^a)$ the anti-hermitian 
generators of the fermion representation of the gauge group.
If this tensor vanishes (i.e.~if one has chosen an {\em anomaly-free}\/
fermion multiplet), the uniqueness and gauge invariance of the theory
can be proved to all orders of perturbation theory.

On the lattice one would like to formulate these 
theories in such a way that

\begin{itemize}
\raggedright
\item[(a)]{one has the right number and type of fermions from the beginning,}
\item[(b)]{locality is preserved,}
\item[(c)]{gauge invariance is unbroken,}
\item[(d)]{the perturbation expansion has the conventional form.}
\end{itemize}

\noindent
The locality of the lattice theory is essential, 
since the universality of the continuum limit depends on it.
Exact gauge invariance is also very impor\-tant
as otherwise the gauge degrees of freedom are not guaranteed 
to decouple from the physical sector of the theory.
Many attempts to put chiral gauge theories on the lattice
in fact failed for this reason \cite{ShamirReview}.
Requirement (d) is included in the list, because perturbation 
theory is currently the only framework where one has
analytical control over the continuum limit
\cite{Reisz,LesHouchesI}.

Until recently it seemed to be impossible to 
fulfil all these conditions simultaneously.
The situation has now changed completely as a result of
the developments following
the rediscovery of the 
Ginsparg--Wilson relation \cite{GinspargWilson}, 
\begin{equation}
  \dirac{5}D+D\dirac{5}=a D\dirac{5}D,
  \label{GWrelation}
\end{equation}
and the construction of gauge-covariant lattice Dirac operators
$D$ satisfying this equation
\cite{PerfectDiracOperator,OverlapDiracOperator}
(here and below $a$ denotes the lattice spacing).

It is striking that two totally different
lines of research produced such Dirac operators at about the same time.
One of them goes under the heading of the perfect-action approach
to asymptotically free lattice field theories \cite{PerfectAction} 
(see ref.~\cite{PerfectActionReview} for a review and further references).
The Dirac operators that are obtained in this framework
provide particular solutions of eq.~(\ref{GWrelation}), with 
good localization properties and small cutoff dependence
\cite{PerfectDiracOperator}.
The other theoretical development started 
with Kaplan's observation \cite{Kaplan}
that the chiral nature
of fermions bound to a 4-dimensional defect in 4+1 dimensions
is preserved on the lattice.
The overlap representation of the fermion determinant
\cite{NarayananNeuberger,RandjbarStrathdee}
derives from this 
and also the 
domain-wall fermion formulation of lattice QCD
\cite{Shamir,FurmanShamir}.
In both cases an effective Dirac operator can be extracted 
\cite{OverlapDiracOperator,KikukawaNoguchi},
which turns out to satisfy the Ginsparg--Wilson relation.

The significance of eq.~(\ref{GWrelation})  
has only been fully appreciated last year.
A key step has been to note that the zero-modes of
any such Dirac operator are chiral and that 
the associated index is a topological invariant which represents
the Chern character on the lattice \cite{IndexTheorem}.
Shortly after this the relation was shown to 
imply an exact chiral symmetry \cite{ExactSymmetry}
with the correct flavour-singlet anomaly
\cite{IndexTheorem,KikukawaYamada,Chiu,ChiuHsieh,ReiszRothe,%
Fujikawa,Suzuki,Adams}.
Left- and right-handed fermions are then easily introduced
\cite{PerfectSplit,OverlapSplit,BoulderReview,AbelianChGT}
and (after a lot more work) this has now led to a formulation of 
abelian chiral gauge theories on the lattice, which
complies with all the basic requirements including
exact gauge invariance \cite{AbelianChGT}. 
The construction is completely general and 
extends to any gauge group \cite{NonAbelianChGT},
but there are still a few loose ends
in the non-abelian case.

In the following we focus on the most recent advances in this field.
Particular attention will be paid 
to the cancellation of the gauge anomaly 
on the lattice. Local cohomology plays an important r\^ole here
and the well-known results on the structure of the anomaly in continuum
chiral gauge theories 
\cite{StoraI,StoraII,Zumino,BaulieuI,BaulieuII,AlvarezGinsparg}
turn out to be very useful at this point
(for a review and an extensive list of references 
see refs.~\cite{EriceLectures,Bertlmann}).
There are also intriguing links 
\cite{KaplanSchmaltz,SuzukiAction,KikukawaAoyama}
to the earlier work of Alvarez-Gaum\'e et al.~\cite{AlvarezEtAl,NielsBohr} 
and of Ball and Osborn \cite{BallOsborn,Ball} relating
the chiral determinant to the $\eta$-invariant of the Dirac operator
in 4+1 dimensions, but this topic will not be addressed here.

\section{WEYL FERMIONS}

Once it had been understood how to preserve chiral symmetry
on the lattice without having to compromise in other ways,
the projection to chiral fermions turned out to 
be straightforward \cite{PerfectSplit,OverlapSplit,BoulderReview,AbelianChGT}.
One begins by choosing any gauge-covariant lattice Dirac operator
$D$ which

\begin{itemize}
\raggedright
\item[(a)]{satisfies the Ginsparg--Wilson relation
           and the hermiticity condition $D^{\dagger}=\dirac{5}D\dirac{5}$,}
\item[(b)]{respects the lattice symmetries,}
\item[(c)]{has the correct behaviour in the free fermion limit,}
\item[(d)]{is local and smoothly dependent on the gauge field.}
\end{itemize}

\noindent
Neuberger's operator \cite{OverlapDiracOperator}
(with the link variables $U(x,\mu)$ replaced by
$R[U(x,\mu)]$ to ensure the correct gauge transformation behaviour)
provides an example of such a lattice Dirac operator.
In this case property (d) is rigorously guaranteed 
if one assumes that the gauge field satisfies
\begin{equation}
  \left\|1-R[U(p)]\right\|<\epsilon
  \label{PlaquetteBound}
\end{equation}
for all plaquettes $p$, where $U(p)$ denotes the product of the link variables
around $p$ and $\epsilon$ any fixed positive number smaller than $\frac{1}{30}$
\cite{Locality}. Usually not all statistically relevant fields 
are of this type, but this can be enforced
through a simple modification
of the Wilson plaquette action \cite{AbelianChGT}.
As far as the continuum limit in the weak coupling phase is concerned,
such actions are perfectly acceptable,
because the bound (\ref{PlaquetteBound})
constrains the gauge field fluctuations at the scale of the cutoff only
and does not violate the locality or
the gauge invariance of the theory.

Chiral fields may now be defined as follows.
One first observes that the operator 
\begin{equation}
  \dirachat=\dirac{5}(1-aD) 
  \label{GammaHat}
\end{equation}
satisfies
\begin{equation}
  (\dirachat)^{\dagger}=\dirachat,
  \quad
  (\dirachat)^2=1,
  \quad
  D\dirachat=-\dirac{5}D.
  \label{PropGammaHat}
\end{equation}
The fermion action
\begin{equation}
  S_{\rm F}=a^4\sum_x\psibar(x)D\psi(x)
  \label{FermionAction}
\end{equation}
thus splits into left- and right-handed parts if the chiral 
projectors for fermion and antifermion fields 
are defined through
\begin{equation}
  \hat{P}_{\pm}=\frac{1}{2}(1\pm\dirachat),
  \quad
  P_{\pm}=\frac{1}{2}(1\pm\dirac{5}),
\end{equation}
respectively. In particular, by imposing the constraints
\begin{equation}
  \hat{P}_{-}\psi=\psi,
  \quad
  \psibar P_{+}=\psibar,
\end{equation}
the right-handed components are eliminated 
and one obtains a classical lattice theory where
a multiplet of left-handed Weyl fermions couples to the 
gauge field in a consistent way. 

\section{FERMION MEASURE}

To define the theory beyond the classical level
one also needs to specify the functional integration measure.
For the gauge field one can take the standard measure,
but the definition of the measure for left-handed fermions
turns out to be non-trivial,
because the projector $\hat{P}_{-}$ depends on the gauge field.

To make this clear, 
let us suppose that $v_j(x)$, $j=1,2,3,\ldots\,$, is 
a basis of complex-valued lattice Dirac fields such that
\begin{equation}
  \hat{P}_{-}v_j=v_j,
  \quad
  (v_k,v_j)=\delta_{kj},
\end{equation}
the bracket being the obvious scalar product for such fields.
The quantum field may then be expanded according to
\begin{equation}
  \psi(x)=\sum_j\,v_j(x)c_j,
\end{equation}
where the coefficients $c_j$ generate a Grassmann algebra.
They represent the independent degrees of freedom of the field
and an integration measure for left-handed fermion
fields is thus given by
\begin{equation}
  \rmD[\kern0.5pt\psi\kern0.5pt]=\prod_{j}\,\rmd c_j.
\end{equation}
Evidently if we pass to a different basis
\begin{equation}
  \tilde{v}_j(x)=\sum_l\,v_l(x)(\trans^{-1})_{lj},
  \quad
  \tilde{c}_j=\sum_l\,\trans_{jl}c_l,
\end{equation}
the measure changes by the factor $\det\trans$,
which is a pure phase factor since $\trans$ is unitary.

It follows from this that
any two bases $v_j$ and $\tilde{v}_j$,
which are related to each other by a unimo\-dular transformation,
define the same measure. Specifying
an integration measure for left-handed fermions thus
amounts to choosing a basis $v_j$ modulo such transformations.
Moreover,
once a particular choice has been made, any other measure is obtained
by multiplication with a phase factor.
An important point to understand
here is that the measure implicitly depends on the gauge 
field through the basis vectors $v_j$.
In general the phase ambiguity is hence gauge-field-dependent too
and does not cancel in expectation values.

In the case of the antifermion fields the subspace of left-handed fields
is independent of the gauge field and one can take the same
orthonormal basis $\bar{v}_k(x)$ for all gauge fields.
The ambiguity in the integration measure
\begin{equation}
  \rmD[\kern0.5pt\psibar\kern0.5pt]=\prod_k\,\rmd \bar{c}_k,
  \quad
  \psibar(x)=\sum_k\,\bar{c}_k\bar{v}_k(x),
\end{equation}
is then only a constant phase factor.

The fermion integral of any product $\cal O$ of the basic
fields may now be defined through
\begin{equation}
  \langle{\cal O}\rangle_{\rm F}=
  \int\rmD[\kern0.5pt\psi\kern0.5pt]\rmD[\kern0.5pt\psibar\kern0.5pt]
  \,{\cal O}\,\rme^{-S_{\rm F}}.
\end{equation}
In particular, if $D$ has no zero-modes, the fermion two-point function
is given by
\begin{equation}
  \langle\psi(x)\psibar(y)\rangle_{\rm F}=
  \langle 1\rangle_{\rm F}\times
  \hat{P}_{-}S(x,y)P_{+},
  \label{Propagator}
\end{equation}
where $S(x,y)$ denotes the Green function of $D$. 
The integral of any other product of fields is then obtained 
by applying Wick's theorem. Note that only the left-handed
components of the fermion field propagate (as it should be).

To obtain the full correlation functions one finally
has to integrate over the gauge field variables as usual.
The definition of the theory is then complete apart from 
the fact that the phase of the fermion measure has not been fixed. 
In the following we shall
almost exclusively be concerned with this problem.
We first determine
which properties the measure must have in order to comply with
the basic principles and then address the question of whether such measures 
exist.

\section{LOCALITY \& GAUGE INVARIANCE}

The locality properties of a euclidean field theory are usually obvious
from the classical action. 
Although the action and the chiral projectors are local, 
the situation is more complicated here,
because the fermion measure is not a simple product of lo\-cal 
measures.

To gain some insight into this problem, let us consider
the fermion partition function
\begin{equation}
  \langle 1\rangle_{\rm F}=\det M,
  \quad
  M_{kj}\equiv a^4\sum_{x}\,\bar{v}_k(x)Dv_j(x),
  \label{detM}
\end{equation}
in the vacuum sector.
If we vary 
the link variables $U(x,\mu)$ in some direc\-tion $\eta_{\mu}(x)$,
\begin{equation}
  \delta_{\eta}U(x,\mu)=a\eta_{\mu}(x)U(x,\mu),
  \label{FieldVariation}
\end{equation}
the effective action changes according to
\begin{equation}
  \delta_{\eta}\ln\det M=\Tr\bigl\{
  \delta_{\eta}D\hat{P}_{-}D^{-1}P_{+}\bigr\}-i\L_{\eta}.
  \label{EffectiveAction}
\end{equation}
The first term in this equation is the naively expected one while
the second, 
\begin{equation}
  \L_{\eta}=i\sum_j\,(v_j,\delta_{\eta}v_j),
  \label{MeasureTerm}
\end{equation}
arises from the 
fermion integration measure. 
$\L_{\eta}$ is linear
in the variation $\eta_{\mu}(x)$ and a current 
$j_{\mu}(x)$ may thus be introduced through
\begin{equation}
  \L_{\eta}=
  a^4\sum_x\,\eta^a_{\mu}(x)j^a_{\mu}(x).
  \label{Current}
\end{equation}
Note that $j_{\mu}(x)$ is a
well-defined expression in the gauge field once
the fermion integration measure has been specified.

On the level of the effective action,
the implicit dependence of the measure on the 
gauge field thus generates additional 
gauge field vertices.
The current $j_{\mu}(x)$ also appears on the right-hand side of 
the field equation of the gauge field as a second term
besides the current derived from the classical action.
To preserve the locality of the theory the fermion measure
should hence be chosen so that the associated 
current $j_{\mu}(x)$ is a local field.
In eq.~(\ref{EffectiveAction}) the measure term then assumes the 
form of a local counterterm and 
the field equations become
relations between local operator insertions as usual.

For gauge variations
$\eta_{\mu}(x)=-\nab{\mu}\omega(x)$, the change
of the effective action is given by
\begin{eqnarray}
  &&\hspace{-1.9em}\delta_{\eta}\ln\det M= \nonumber\\[1ex]
  &&\hspace{-1.9em}\quad
    ia^4\sum_x\,\omega^a(x)
  \bigl\{\anomaly^a(x)-[\nabstar{\mu}j_{\mu}]^a(x)\bigr\},\\[1ex]
  &&\hspace{-1.9em}\anomaly^a(x)={ia\over2}
  \kern1.0pt\tr\{\dirac{5}R(T^a)D(x,x)\}.
  \label{Anomaly}
\end{eqnarray}
In these equations $\nab{\mu}$ and $\nabstar{\mu}$ denote
the gauge-covariant forward and backward lattice derivatives and
$D(x,y)$ the kernel of the Dirac operator in position space.

From the definition (\ref{Anomaly}) and the properties
of the Dirac operator it is obvious that 
$\anomaly(x)$ is a gauge-covariant local expression in the gauge field.
It can easily be worked out in the classical continuum limit,
where the link variables are assumed to be given by a
smooth background gauge potential $A_{\mu}(x)$ through
the usual path-ordered exponentials.
As a result one obtains
\begin{equation}
  \anomaly^a(x)=c_1d^{abc}_R\epsilon_{\mu\nu\rho\sigma}
  F^b_{\mu\nu}(x)F^c_{\rho\sigma}(x)+\rmO(a),
\end{equation}
with $c_1=-1/128\pi^2$
\cite{KikukawaYamada,Chiu,ChiuHsieh,ReiszRothe,Fujikawa,Suzuki,Adams},
and $\anomaly(x)$ thus represents the {\em covariant anomaly}\/ 
on the lattice.

Now if we require that the gauge symmetry be unbroken,
the fermion determinant should be gauge-invariant and from 
the above we then conclude that 
\begin{equation}
  [\nabstar{\mu}j_{\mu}]^a(x)=\anomaly^a(x).
  \label{GaugeInvariance}
\end{equation}
Moreover, recalling eq.~(\ref{EffectiveAction}), it is obvious
that the current $j_{\mu}(x)$ has to be a gauge-covariant 
expression in the gauge field.
The converse is also true, i.e.~any 
fermion integration measure that yields
a current with these properties 
preserves the gauge symmetry
\cite{AbelianChGT,NonAbelianChGT}.

As is well known one cannot have both locality and 
gauge invariance, unless the anomaly cancellation condition
\begin{equation}
  d_{R}^{abc}=0
\end{equation}
is fulfilled. In the present framework this 
may be shown by noting that 
eq.~(\ref{GaugeInvariance}) has no solution in the classical
continuum limit if the anomaly does not vanish at $a=0$.

\section{INTEGRABILITY CONDITION}

So far we have assumed that the current $j_{\mu}(x)$ is obtained 
from a given fermion integration measure through 
eqs.~(\ref{MeasureTerm}),(\ref{Current}). 
The question may now be asked whether one could also
start with a prescribed current and
derive the measure from it.
As explained below this is indeed the case
if the current satisfies a certain integrability condition.
The construction of the lattice theory
is thus significantly simplified, because the measure
(which is a relatively complicated object)
no longer needs to be specified explicitly.

To derive the integrability condition, let us consider a 
smooth curve
\begin{equation}
  U_t(x,\mu),\quad 0\leq t\leq 1,
  \label{Curve}
\end{equation}
in field space. 
The change of the effective action along this curve is given by
\begin{eqnarray}
  &&\hspace{-1.9em}
  \partial_t\ln\det M=\Tr\bigl\{\partial_tD
  \hat{P}_{-}D^{-1}P_{+}\bigr\}-i\L_{\eta},\\[1ex]
  &&\hspace{-1.9em}
  a\eta_{\mu}(x)=\partial_t U_t(x,\mu)U_t(x,\mu)^{-1}
  \label{eta_t}
\end{eqnarray}
(for simplicity the $t$-dependence of
$\eta_{\mu}(x)$ is suppressed).
At $t=1$ the solution of this equation yields
\cite{NonAbelianChGT}
\begin{eqnarray}
  &&\hspace{-1.9em}
  \det M \det {M_0}^{\kern-1pt\dagger}= \nonumber\\[1ex]
  &&\hspace{-1.9em}
  \quad\det\bigl\{1-P_{+}+P_{+}DQ_1{D_0}^{\kern-1pt\dagger}\bigr\}W^{-1},
  \label{FermionDeterminant}
\end{eqnarray}
where 
\begin{equation}
  W=\exp\biggl\{i\int_0^1\rmd t\,\L_{\eta}\biggr\}
  \label{WilsonLine}
\end{equation}
is the total change of phase of the fermion measure
and the operator $Q_t$ is defined through
\begin{equation}
  \partial_tQ_t=\left[\partial_t P_t,P_t\right]Q_t,
  \quad P_t=\hat{P}_{-}|_{U=U_t},
\end{equation}
with initial value $Q_0=1$. $Q_t$ is unitary
and satisfies $P_tQ_t=Q_tP_0$, i.e.~it maps left-handed
fields at $U=U_0$ to left-handed fields at $U=U_t$.

We now note
that the left-hand side of eq.~(\ref{FermionDeterminant}) 
only depends on the end-points of the chosen path in field space.
On the other side of the equation,
a closer examination shows this to be 
a consequence of the fact that 
\begin{equation}
  W=\det\left\{1-P_0+P_0Q_1\right\}
  \label{GIC}
\end{equation}
for all {\em closed curves}.
The important point to understand here is that this identity
holds whenever the current $j_{\mu}(x)$ is obtained 
from an underlying fermion measure in the way we
have described. In other words, if we start from an 
arbitrary current and define the phase factors $W$ through
eqs.~(\ref{Current}),(\ref{eta_t}),(\ref{WilsonLine}),
the validity of eq.~(\ref{GIC}) 
is a necessary condition for the current to be associated 
with a fermion measure. 

As it turns out,
this is in fact also a sufficient condition \cite{NonAbelianChGT}. 
Moreover the current determines the
measure uniquely
up to a constant phase factor in each topological sector.
The construction of the lattice theory
is thus reduced to the task of finding a current $j_{\mu}(x)$
that

\begin{itemize}
\raggedright
\item[(a)]{is a gauge-covariant local expression in the gauge field,}
\item[(b)]{transforms like an axial vector current under the lattice
           symmetries,}
\item[(c)]{satisfies the anomalous conservation law~(\ref{GaugeInvariance}),}
\item[(d)]{fulfils the integrability condition~(\ref{GIC}).}
\end{itemize}

\noindent
Once this is achieved, the lattice theory
is completely specified and guaranteed to
comply with the basic principles listed earlier.

At first sight the integrability condition
looks quite inaccessible, since it involves the
determinant of a complicated operator. 
In its differential form,
\begin{eqnarray}
  &&\hspace{-1.9em}
  \delta_{\eta}\L_{\zeta}-\delta_{\eta}\L_{\zeta}+a\L_{[\eta,\zeta]}=
  \nonumber\\[1ex]
  &&\hspace{-1.9em}
  \hspace{6em} i\kern1pt\Tr\bigl\{\hat{P}_{-}
  [\delta_{\eta}\hat{P}_{-},\delta_{\zeta}\hat{P}_{-}]
  \bigr\},
  \label{LIC}
\end{eqnarray}
the equation is, however, much more tractable.
It can be shown, for example, that 
the right-hand side of eq.~(\ref{LIC}) is proportional to 
\begin{equation}
  \int\rmd^4x\,d^{abc}_R\epsilon_{\mu\nu\rho\sigma}\kern1pt
  \eta^a_{\mu}(x)\zeta^b_{\nu}(x)F^c_{\rho\sigma}(x)
\end{equation}
in the classical continuum limit. This expression vanishes
if the fermion multiplet is anomaly-free, and setting
$j_{\mu}(x)=0$ in this limit then fulfils both
the requirement of gauge 
invariance and the integrability condition in its differential form.
As far as the classical continuum limit goes, this completes
the definition of the lattice theory.
In particular, in the formula (\ref{FermionDeterminant})
for the fermion determinant, the phase factor $W$ 
is equal to $1$ up to terms vanishing proportionally to $a$.

A last comment which should be made here is that 
our discussion of the fermion determinant is in
many respects similar to
Leutwyler's approach in the continuum theory \cite{Leutwyler},
where one first applies a finite-part prescription to 
the variation of the determinant and then adds a 
local counter\-term to restore the integrability of the expression.
The measure term $\L_{\eta}$ 
effectively plays the r\^ole of this counterterm and
eq.~(\ref{LIC}) may actually be derived directly from
eq.~(\ref{EffectiveAction}) by noting
that the right-hand side has to be integrable.

\section{U(1) THEORIES}

At this point we still need to prove that 
there exists a current $j_{\mu}(x)$ with the required properties
if the fermion multiplet is anomaly-free.
It suffices to construct the current on the subset of all
gauge fields satisfying the bound (\ref{PlaquetteBound}),
since only these contribute to the functional integral.
While this is technically helpful, a definitive answer to the 
question has so far only been given for abelian gauge groups
\cite{AbelianChGT}.

The simplest case to consider is a theory with
$N$ left-handed fermions coupled to a U(1) gauge field.
Without loss the fermion representation of the 
gauge group may be taken to be diagonal,
\begin{equation}
  R[\Lambda]_{\alpha\beta}=\Lambda^{\rme_{\alpha}}\delta_{\alpha\beta},
  \quad \Lambda\in\rmU(1),
\end{equation}
where the indices label the fermion flavours and the integers 
$\rme_{\alpha}$ denote their charges.
The ano\-maly cancellation condition then reads
\begin{equation}
  \sum_{\alpha=1}^N\rme_{\alpha}^3=0
  \label{AnomalyFree}
\end{equation}
and an interesting example of an anomaly-free charge assignment is thus
\begin{equation}
  \rme_1=\ldots=\rme_8=1,\quad \rme_9=-2.
  \label{ExampleMultiplet}
\end{equation}
The main theorem established in 
ref.~\cite{AbelianChGT} applies to the theory in infinite volume 
and asserts that a current satisfying all conditions
(a)--(d) listed above exists if eq.~(\ref{AnomalyFree}) holds.
The proof of the theorem is constructive and one ends up with a
complicated but well-defined expression for the current.

In finite volume with periodic boundary conditions,
the space of gauge orbits divides into 
topological sectors with non-contractible closed loops.
Global obstructions can then arise,
but for most charge assignments,
including the multiplet (\ref{ExampleMultiplet}) and all representations
with only even charges, the theorem extends to 
all topological sectors in finite volume \cite{AbelianChGT}.

A more detailed description of this result and its derivation 
is beyond the scope of this review. 
Instead we now briefly discuss how 
the anomaly cancellation works out on the lattice.
In the U(1) theories considered in this section,
the anomaly (\ref{Anomaly}) assumes the form
\begin{equation}
  {\cal A}(x)=-\frac{1}{2}a\kern1pt\tr\{\dirac{5}TD(x,x)\},
  \label{AbelianAnomaly}
\end{equation}
where $T$ is the charge matrix
$T_{\alpha\beta}=\rme_{\alpha}\delta_{\alpha\beta}$.
The anomaly
is thus a gauge-invariant local expression in the gauge field.
A less obvious property is that 
\begin{equation}
  a^4\sum_x\delta{\cal A}(x)=0
  \label{TopField}
\end{equation}
for any local variation $\delta U(x,\mu)$ of the gauge field.
To prove this one first notes that the left-hand side of eq.~(\ref{TopField})
is proportional to $\Tr\{T\delta\dirachat\}$.
Using the identities
\begin{equation}
  (\dirachat)^2=1,\quad\{\dirachat,\delta\dirachat\}=0,
  \quad [\dirachat,T]=0,
\end{equation}
the trace is then easily seen to vanish.

Equation (\ref{TopField}) says that 
the abelian anomaly is a {\em topological field},
i.e.~it has all the characteristic 
properties of a topological density.
Modulo divergence terms 
(which are topologically uninteresting)
there are usually not many
fields of this type and 
the form of the anomaly is thus strongly constrained.
To make this explicit, we first note that
the Chern polynomial
\begin{eqnarray}
  &&\hspace{-1.9em}
  c(x)=\alpha+\beta_{\mu\nu}F_{\mu\nu}(x)+ \nonumber\\[1ex]
  &&\hspace{-1.9em}\hspace{2em}
  \gamma\epsilon_{\mu\nu\rho\sigma}
  F_{\mu\nu}(x)F_{\rho\sigma}(x+a\hat{\mu}+a\hat{\nu})
  \label{ChernPolynomial}
\end{eqnarray}
provides a simple example of a non-trivial topological field.
The field tensor $F_{\mu\nu}(x)$, which appears here,
is defined as usual from the plaquette loops
and $\hat{\mu}$ denotes the unit vector in direction~$\mu$
(the displacement of the coordinates in the last term accounts
for the fact that the Leibniz rule is modified on the lattice).
The crucial observation is now that up to divergence
terms there are actually no further topological fields
\cite{AbelianCohomologyI,AbelianCohomologyII}.
In other words, any topological field $q(x)$
constructed from a U(1) lattice gauge field
is of the form
\begin{equation}
  q(x)=c(x)+\drvstar{\mu}k_{\mu}(x),
\end{equation}
where $k_{\mu}(x)$ is a {\em gauge-invariant local}\/ current.
In particular, the anomaly ${\cal A}(x)$ can be written in this
way, with some
constants $\alpha$, $\beta_{\mu\nu}$ and $\gamma$.

It is easy to see that $\alpha$ and $\beta_{\mu\nu}$
have to vanish, because the anomaly transforms as a 
pseudo-scalar under the lattice symmetries. 
Concerning the coefficient $\gamma$ we note that
the anomaly is a sum of terms, one for each fermion flavour.
Since the field tensor scales with the charge and 
since there is another power of the charge coming from the 
charge matrix $T$ in eq.~(\ref{AbelianAnomaly}), we have
\begin{equation} 
  \gamma\propto\sum_{\alpha=1}^N\rme_{\alpha}^3.
\end{equation}
The topologically non-trivial part of the anomaly
thus cancels if the fermion multiplet is anomaly-free
and the condition for exact gauge invariance, eq.~(\ref{GaugeInvariance}),
then reduces to 
\begin{equation}
  \drvstar{\mu}j_{\mu}(x)=\drvstar{\mu}k_{\mu}(x).
\end{equation}
Although more work is required to actually 
show this \cite{AbelianChGT},
it should now be quite plausible
that this relation can be satisfied. At least the topological
obstruction represented by the anomaly has completely
disappeared at this stage.

\section{NON-ABELIAN THEORIES}

If the gauge group is not abelian,
the anomaly has a more complicated structure and 
does not seem to have any obvious topological properties.
In the continuum limit
the anomaly is, however, known
to be closely related to
the Chern character in 4+2 dimensions.
Algebraically the link is provided by the descent equations
\cite{StoraI,StoraII,Zumino,BaulieuI,BaulieuII}
and there is also an associated 
topological interpretation of the anomaly
\cite{AlvarezGinsparg}.
Somewhat surprisingly
the relation persists on the lattice and
the exact cancellation of the anomaly is then again 
mapped to the problem of classifying topological fields,
although this time in 4+2 dimensions \cite{NonAbelianChGT}.

We now describe this result
in outline and
begin by considering gauge fields 
\begin{equation}
  U(z,\mu),\;A_t(z),\;A_s(z),
  \quad
  z=(x,t,s),
\end{equation}
on $\LxRtwo$,
the continuous extra coordinates being $t$ and $s$.
The gauge potentials in these directions take values
in the Lie algebra of the gauge group as usual. They 
allow one to define
gauge-covariant derivatives such as 
\begin{equation}
  \D_t\hat{P}_{-}=\partial_t\hat{P}_{-}+[R(A_t),\hat{P}_{-}].
\end{equation}
In particular, the field
\begin{eqnarray}
  &&\hspace{-1.9em}
  q(z)= 
  \frac{1}{2}\,\Im\kern1pt\tr\,\Bigl\{
  \bigl[\dirachat[\D_t\hat{P}_{-},\D_s\hat{P}_{-}]+
  \nonumber\\[1ex]
  &&\hspace{-1.9em}
  \hspace{8em}
  R(F_{ts})\dirachat\bigr](x,x)\Bigr\}
  \label{q-field}
\end{eqnarray}
(where $[\ldots](x,y)$ denotes the kernel in position space 
representing the operator
enclosed in the square bracket) is invariant
under arbitrary gauge transformations in 4+2 dimensions.
$q(z)$ is also a local field and it can be shown to be topological
in the sense explained above, viz.
\begin{equation}
  a^4\sum_x\int\rmd t\,\rmd s\,
  \delta q(z)=0
\end{equation}
for all local deformations of the gauge field.

The importance of all this is now made clear
by the following

\vskip1ex
\noindent
{\sl
{\bf Theorem.} The field $q(z)$ is topologically trivial
if and only if there exists a gauge-covariant local current $j_{\mu}(x)$ 
satisfying the integrability condition in its differential form,
eq.~(\ref{LIC}),
and the anomalous conservation law (\ref{GaugeInvariance}).}

\vskip1ex
\noindent
To establish the exact cancellation of the anomaly on the lattice
we thus need to show that $q(z)$ is topologically trivial.
Presumably there are just a few non-trivial topological fields
on $\LxRtwo$.
The field $q(z)$ is a linear combination
of these plus a divergence term and one then has to prove that 
the coefficients of the non-trivial fields vanish if 
the fermion multiplet is anomaly-free. 

The classification of topological fields modulo divergence terms
is a particular case of a {\em local cohomology}\/ problem,
a subject that has received a lot of attention in 
continuum field theory.
In particular,
using the descent equations \cite{StoraI,StoraII,Zumino,BaulieuI,BaulieuII},
a general theorem has been established,
for any gauge group and in any dimension,
which states that in the absence of matter fields the Chern monomials
are the only non-trivial topological fields
\cite{BrandtEtAl,DuboisVioletteEtAl,Dragon}.

Although this result is for the continuum theory, it allows us
to prove the anomaly cancellation on the lattice to
all orders of an expansion in powers of $a$ around
the classical continuum limit.
As explained before, the limit is approached by
assuming the link variables $U(z,\mu)$ to be given by a
smooth background gauge potential $A_{\mu}(z)$.
For $a\to0$ the asymptotic expansion
\begin{equation}
  q(z)=
  \sum_{k=0}^{\infty}a^{k-6}{\cal O}_k(z)
  \label{a-expansion}
\end{equation}
is then obtained, ${\cal O}_k(z)$ being polynomials 
of dimension $k$ in the potentials
$A_{\mu}(z)$, $A_t(z)$, $A_s(z)$ and their derivatives.
The first few terms are not difficult to work out 
and one finds that
\begin{eqnarray}
  &&\hspace{-1.9em}
  q(z)=\frac{1}{6}c_1d^{abc}_R\epsilon_{\mu_1\ldots\mu_6}
  \nonumber\\[1ex]
  &&\hspace{-1.9em}
  \hspace{2em}
  \times F^a_{\mu_1\mu_2}(z)F^b_{\mu_3\mu_4}(z)F^c_{\mu_5\mu_6}(z)
  +\rmO(a),
\end{eqnarray}
where the obvious notations are being used,
with space-time indices running from 0 to 5.
In the continuum limit $q(z)$ is thus proportional to 
the Chern character in 4+2 dimensions \cite{Gilkey}. 
The connection between
the anomaly and the Chern character is thus made explicit and 
the equation also shows
that all fields ${\cal O}_k(z)$ with dimension $k\leq6$ vanish
if the fermion multiplet is anomaly-free.

We now note that the higher-order terms
have to be gauge-invariant and topological, because 
$q(z)$ has these properties.
The classification theorem quoted above then
implies that they are all equal
to a sum of Chern monomials plus a divergence term.
Since there are no Chern monomials with scale dimension greater
than the space-time dimension, 
we thus conclude that $q(z)$ is topologically trivial to 
all orders in $a$ if the anomaly cancels at $a=0$.

Evidently it is not certain that this result
extends to any fixed value of the lattice spacing,
but it would be a surprise if it did not.
What is lacking at present is a classification theorem
for topological fields on $\LxRtwo$.
Presumably the non-trivial classes match
with the non-trivial classes in the continuum theory,
and the exact cancellation of the anomaly on the lattice
would then be obvious from our discussion above.

\section{CONCLUDING REMARKS}

The starting point in this review has been 
the Ginsparg--Wilson relation, which implies an exact
chiral symmetry of the fermion action and thus allows one to introduce
Weyl fermions in a sensible way.
Another consequence of this relation is that 
the gauge anomaly on the lattice 
descends from a topological field in 4+2 dimensions.
For the exact cancellation of the anomaly, this 
property is essential as otherwise
there is no reason for the
lattice corrections to the anomaly to cancel
if the fermion representation is anomaly-free.

An important topic, which we did not discuss,
are global anomalies \cite{Witten,ElitzurNair}.
Such anomalies can arise 
if there are non-contractible loops in field space, because
the integrability condition (\ref{GIC})
is then a slightly stronger constraint than its differential form.
The latter implies
\begin{equation}
  W=h\det\left\{1-P_0+P_0Q_1\right\}
\end{equation}
for all closed curves in field space, where $h$ is a constant phase factor
depending on the homotopy class of the curve.
Evidently $h=1$ for all contrac\-tible loops,
but in general this need not be so. 

For illustration let us consider a Weyl fermion 
coupled to an SU(2) gauge field in the fundamental representation.
The anomaly and the right-hand
side of eq.~(\ref{LIC}) vanish
in this case and setting $j_{\mu}(x)=0$ thus 
appears to fulfil all conditions.
As has recently been shown by B\"ar and Campos
\cite{BaerCampos},
there are, however, closed loops in field space
for which 
\begin{equation}
  \det\left\{1-P_0+P_0Q_1\right\}=-1.
\end{equation}
Moreover they explain in detail how this relates 
to Witten's original derivation of the anomaly and
the conclusion is then that there is no
acceptable fermion measure in this theory.

While this leaves little doubt that
the known global anomalies can all be reproduced on the lattice,
it is not obvious that   
such anomalies are absent in those theories 
where they are not expected to occur.
So far this has only been shown for abelian gauge groups
\cite{AbelianChGT}.

Although only purely left-handed models have been considered here,
no additional difficulties are expected to arise
when more general multiplets of Weyl fermions
and Higgs fields are included.
In particular, one may now envisage to study
non-leptonic weak decays using an exactly gauge-invariant lattice
formulation of the Standard Model.
To check the $\Delta I=1/2$ rule, for example,
an interesting option in such an approach is
to set the mass of the $W$ boson to unphysically low values.
The operator product expansion is then not needed and 
the hard renormalization problems that go with it are thus avoided
(for a recent discussion of this issue and a related proposal
see ref.~\cite{DawsonEtAl}).

\vskip1ex
I am indebted to 
Oliver B\"ar, Isabel Campos,
Peter Hasenfratz, Pilar Hern\'andez, Karl Jansen, Ferenc Niedermayer,
Raymond Stora and Peter Weisz
for correspondence and many 
helpful discussions on the topics covered in this review.

\end{document}